\documentclass[final,5p,times,twocolumn,authoryear]{elsarticle}

\usepackage{amssymb}
\usepackage{amsthm}
\usepackage{bbm}
\usepackage{amsfonts, float, mathtools}

\newcommand{\caA}{{\mathcal A}}
\newcommand{\cA}{{\mathcal A}}
\newcommand{\cB}{{\mathcal B}}

\newcommand{\cG}{{\mathcal G}}
\newcommand{\cH}{{\mathcal H}}

\newcommand{\cP}{{\mathcal P}}

\newcommand{\cS}{{\mathcal S}}

\newcommand{\bC}{{\mathbb C}}

\newcommand{\bR}{{\mathbb R}}

\newcommand{\bZ}{{\mathbb Z}}

\newcommand{\idty}{\mathbbm{1}}
\newcommand{\tr}{\mathrm{tr}}
\newcommand{\Tr}{\mathrm{Tr}}
\newcommand{\gap}{\mathrm{gap}}

\newcommand{\diam}{\mathrm{diam}}
\newcommand{\Span}{\mathrm{span}}
\newcommand{\ran}{{\rm ran}}

\newcommand{\spec}{\mathop{\rm spec}}

\newcommand{\loc}{{\rm loc}}
\newcommand{\supp}{{\rm supp}}

\newcommand{\braket}[2]{\left\langle #1 , #2\right\rangle}

\newcommand{\ket}[1]{\vert #1 \rangle}

\begin{document}

\begin{frontmatter}



\title{Quantum Spin Systems}


\author[first,second]{Amanda Young}
            \ead{ayoung86@illinois.edu}
\affiliation[first]{organization={Department of Mathematics, University of Illinois Urbana-Champaign},
            city={Urbana},
            state={IL},
            country={USA}}
\affiliation[second]{organization={Munich Center for Quantum Science and Technology and Center for Mathematics, TU Munich},
            city={Garching},
            country={Germany}}

\begin{abstract}
This work provides an overview of gapped quantum spin systems, including concepts, techniques, properties, and results. The basic framework and objects of interest for quantum spin systems are introduced, and the main ideas behind methods for proving spectral gaps for frustration-free models are outlined. After reviewing recent progress on several spectral gap conjectures, we discuss quasi-locality of the Heisenberg dynamics and its utility in proving properties of gapped quantum spin systems. Lieb-Robinson bounds have played a central role in establishing exponential decay of ground state correlations, an area law for one-dimensional systems, a many-body adiabatic theorem, and spectral gap stability. They also aided in the development of the quasi-adiabatic continuation, which is a useful for investigating gapped ground state phases, both of which are also discussed.
\end{abstract}


\end{frontmatter}

\tableofcontents



\section{Introduction}
\label{introduction}
 Quantum spin systems are mathematical models that are frequently used to represent physical systems in both condensed matter physics and quantum information theory, including models for magnetism and quantum circuits. They are comprised of a countable number of sites, most commonly the vertices of a lattice, and each site has a finite-dimensional Hilbert space of states. For example, the Hilbert space at each site might represent a particle of spin-$s$ fixed at that vertex of a lattice.

 The main goal of analyzing quantum spin models is to apply the standards of mathematical rigor to capture their key physical features. One typically starts by studying the composite system associated to a finite subset of sites. The Hamiltonian in this case is a bounded, self-adjoint operator that encodes the interactions between the contained sites, and its spectrum is the set of possible energies. Important in the study of quantum spin models are properties of the spectrum and the associated energy states as the finite volume grows to include all sites, called the thermodynamic limit. This limiting process can be used to obtain a Hamiltonian for the infinite system. One key feature of the infinite volume Hamiltonian with many consequences is whether or not it has a gap in its spectrum above its minimal energy. The existence or non-existence of this gap is also one of the criteria used to classify a model into its quantum phase of matter. Given its significance, this article focuses on topics related to gapped quantum spin models.

Our aim is to give a broad overview on the subject of gapped quantum spin systems, including important concepts, techniques, properties, and results. In Section~\ref{sec:setup}, the basic mathematical framework for quantum spin systems is introduced, including dynamics and states. Uniformly gapped quantum spin models is the focus of Section~\ref{sec:gap_methods}, where methods for proving gaps, how likely it is for a generic quantum spin model to be gapped (in a probabilistic sense), and progress on long-standing gap conjectures is discussed. Properties exhibited by gapped quantum spin models, such as exponential decay of correlations, and key tools for analyzing gapped ground state phases are the topics of Section~\ref{sec:gap_results}. The work concludes with an overview of spectral gap stability in Section~\ref{sec:Stability}.

 \section{Background}\label{sec:background}

 The first quantum spin model was introduced by Heisenberg shortly after it was discovered that magnetism was a quantum mechanical effect. Since then, investigations of the Heisenberg spin chain have lead to a wealth of progress. One of the earliest and most well-known results is Bethe's solutions for the eigenvalues and eigenvectors for the antiferromagnetic spin-1/2 Heisenberg chain. This showed that the model does not have a spectral gap above its ground state energy, and his approach, known today as the Bethe Ansatz, is applicable to other exactly solvable models. In addition, the Lieb-Schultz-Mattis (LMS) Theorem, which established a set of conditions for when a quantum spin chain cannot have a unique ground state and a positive gap, also stemmed from analyzing the spin-$1/2$ model \citep{Lieb1961}. Its proof is based on a variational calculation, and was the first result to illustrate that one does not need all of the energy states to prove that a model is gapless.

In condensed matter physics, new exotic phases of matter emerge at low temperatures, and whether there is a spectral gap above the ground state energy is one of the main quantities of interest. In the case of $SU(2)$-invariant, antiferromagnetic spin-$s$ chains, Haldane's seminal work predicted a collection of phases in interaction space each of which is classified by whether spin is integral or half-integral \citep{Haldane1983b,Haldane1983a}. Namely, he argued that there exists a gapless phase with a unique ground state that exhibits power-law decaying correlations in the case of half integral $s$, and a gapped phase with a unique ground state with exponentially decaying correlations for integral $s$. He predicted these properties by studying the limit of large $s$, but nevertheless conjectured the same phase existed for small values of the spin as well. Given Bethe's work, the prediction for half integral $s$ was not so surprising, and was quickly verified \citep{Affleck1986}. However, the conjecture for integer $s$, and in particular $s=1$, was not expected. One of the first applications of the density matrix renormalization group (DMRG) method provided numerical evidence supporting that the spin-$1$ Heisenberg model satisfies these properties, but a rigorous proof has still not been found. The phase for $s=1$ is now known as the Haldane phase. 

Inspired by the integral spin conjecture, Affleck, Kennedy, Lieb and Tasaki (AKLT) proposed an $SU(2)$-invariant, antiferromagnetic spin-$1$ chain and proved that this model satisfied the properties of the Haldane phase \citep{Affleck1988}. Much like the Heisenberg model, the AKLT model has been the catalyst for numerous advances in mathematical physics. Their ground states are the preeminent example of Tensor Network States (TNS), and are an example of a universal quantum computation resource. Moreover, all general methods for proving nonvanishing spectral gaps can be traced back to the AKLT model and its analysis. It is also from analyzing this model that it is now understood that the Haldane phase is a non-trivial symmetry protected topological phase.

In addition to understanding the behavior of the spectrum, properties related to the evolution of a quantum spin model are also key for studying quantum spin systems. Even though quantum many-body theory is non-relativistic, Lieb and Robinson showed that the support of any local observable evolved under the Heisenberg dynamics generated by a finite-range quantum spin interaction effectively spreads linearly with time \citep{Lieb1972}. The importance and power of this observation was first appreciated by Hastings, who harnessed Lieb-Robinson bounds to prove a multi-dimensional version the of LSM Theorem. Since then, a plethora of results proving or using quasi-locality of the dynamics have been produced, many of which are aimed at establishing properties of gapped ground state phases. As such, quasi-locality bounds will be specifically highlighted in this discussion of rigorous result related to gapped quantum spin models.

\section{Mathematical foundations and examples}
\label{sec:setup}

 For simplicity, we restrict our focus to quantum spin systems where the spatial degrees of freedom are given by the sites of a lattice $\Gamma$, e.g. $\Gamma=\bZ^D$ for some integer $D\geq 1$, and assume that each site has the same space of states. Thus, there is integer $n\geq 1$ so that the Hilbert space $\cH_{\{x\}}=\bC^n$ represents the state of any site $x\in\Gamma$. For each site, the set of $n\times n$ complex-valued matrices that acts on $\cH_{\{x\}}$ gives the single-site algebra of observables, denoted $\cA_{\{x\}}=M_n(\bC)$. 

For example, if a particle of spin-$s$, $s\in \{0,\frac{1}{2}, 1, \frac{3}{2},\ldots\}$, is fixed at each $x\in\Gamma$, then the state at each site is determined by its spin component and $n=2s+1.$ The spin matrices $S^1, \, S^2,\, S^3\in M_{n}(\bC)$, generate the unique (up to isomorphism) $n$-dimensional irreducible representation of the Lie algebra $\mathfrak{su}(2)$, and are of particular interest for a spin-$s$ quantum spin model. These are hermitian matrices, and their set of eigenvalues are $\{s, s-1, s-2, \ldots, -s\}$. The most common representation of these matrices has the canonical basis of $\bC^n$ as an orthonormal basis of eigenvectors of $S^3$. The other two spin matrices are then defined follows. Let $\ket{j}$ denote the eigenvector of $S^3$ with eigenvalue $j$. The raising operator $S^+\in M_n(\bC)$ is defined on this basis by $S^+\ket{s}=0$ and
\[
S^{+}\ket{j} = \sqrt{s(s+1)-j(j+1)}\ket{j+1}, \;\; -s \leq j \leq s-1.
\]
Then, $S^{1}=(S^+ +S^-)/2$ and $S^{2}=(S^+-S^-)/2i$ where $S^- = (S^+)^*$ is the lowering operator.

Quantum spin models are first defined on finite subsets of the lattice. Denoting by $\cP_\Gamma$ the set of all finite subsets from $\Gamma$, the Hilbert space of states and algebra of observables for the subsystem $\Lambda\in\cP_\Gamma$ are defined by the tensor products of their single site counterparts:
\[\cH_\Lambda = \bigotimes_{x\in\Lambda}\cH_{\{x\}}, \quad \cA_\Lambda = \bigotimes_{x\in\Lambda}\cA_{\{x\}}.\]

The algebra $\cA_\Lambda$ has the structure of a $C^*$-algebra with respect to matrix sum and product, hermitian conjugation, and the operator norm
\[
\|A\| =\sup_{0\neq \psi \in\cH_\Lambda}\frac{\|A\psi\|}{\|\psi\|} \;\; \forall \, A\in\cA_\Lambda \; \text{where}\;\|\psi\|= \sqrt{\braket{\psi}{\psi}}\,.
\]
The identity element is the identity matrix, $\idty_\Lambda\in \cA_\Lambda$, and one can embed $\cA_{\Lambda'}\subseteq \cA_\Lambda$ for any $\Lambda'\subseteq \Lambda$ via the identification
\begin{equation}\label{embedding}
    \cA_{\Lambda'}\ni A\mapsto A\otimes \idty_{\Lambda\setminus\Lambda'}\in \cA_{\Lambda}.
\end{equation}
As a consequence, the algebra of local observables
\[
\cA_{\Gamma}^{\loc} = \bigcup_{\Lambda\in \cP}\cA_\Lambda
\]
is well-defined, and the $C^*$-algebra of quasi-local observables for the infinite system, denoted $\cA_{\Gamma}$, is the closure of $\cA_{\Gamma}^{\loc}$ with respect to the operator norm.

The interaction between the sites $X\in\cP_\Gamma$ is given by a self-adjoint matrix $\Phi(X)^*=\Phi(X)\in\cA_X$. After fixing an interaction, the local Hamiltonian defined by
\begin{equation}\label{local_Ham}
    H_\Lambda = \sum_{X\subseteq \Lambda} \Phi(X) \in \cA_\Lambda, \quad \Lambda\in\cP_\Gamma
\end{equation}
is the sum of all interactions between the sites of $\Lambda.$ It generates the dynamics, and its spectrum is the set of all possible energies of the subsystem. This expression is well-defined by \eqref{embedding}. 

Two of the most important examples of quantum spin models are the Heisenberg and AKLT models on $\Gamma =\bZ$, both of which have an $SU(2)$-invariant interaction. The Heisenberg interaction for the spin-$s$ chain is 
\begin{equation}
    \label{heisenberg}
    \Phi(\{x,x+1\}) = -J \mathbf{S}_x\cdot \mathbf{S}_{x+1}, \quad \forall x\in\bZ
\end{equation}
where, in terms of the $2s+1$-dimensional spin matrices,
\[\mathbf{S}_x\cdot \mathbf{S}_{x+1}= S_x^1\otimes S_{x+1}^1 + S_x^2\otimes S_{x+1}^2 + S_x^3\otimes S_{x+1}^3\]
and $\Phi(X)=0$ for all other $X\in\cP_\bZ.$ The model is ferromagnetic for $J>0$ and antiferromagnetic for $J<0$. The magnitude $|J|$ only scales the energy values of the system, and so one usually sets $|J|=1$. The interaction for the spin-1 AKLT chain is
\begin{equation}\label{aklt}
\Phi^{AKLT}(\{x,x+1\}) = \mathbf{S}_x\cdot \mathbf{S}_{x+1} + \frac{1}{3}\left(\mathbf{S}_x\cdot \mathbf{S}_{x+1}\right)^2,\; \forall x\in\bZ
\end{equation}
and zero for all other $X\in\cP_\bZ.$

Both models are examples of translation invariant, nearest neighbor models. When every lattice translation $\tau$ induces an isomorphism on the local algebras, i.e. $\caA_X \cong \caA_{\tau(X)}$, for all $X\in\cP_\Gamma,$ we say that a model is translation invariant if $\tau(\Phi(X)) = \Phi(\tau(X))$ for all $X\in\cP_\Gamma$ and translations $\tau$. Note that we are slightly abusing notation and using $\tau$ to also represent the algebra isomorphism induced by the translation. A model is nearest neighbor if $\Phi(X)\neq 0$ implies $X=\{x,y\}$ for a pair of neighboring sites, $|x-y|=1.$ More generally, a model is finite range if there exists $R\geq 0$ such that $\Phi(X) \neq 0$ only if $\diam(X) \leq R.$ For simplicity, the metric on $\Gamma$ is taken to be the graph distance throughout this discussion.

The Heisenberg model can be generalized to a nearest neighbor model on $\bZ^d$ by replacing $x+1$ with $y$ in the interaction above for a nearest-neighbor pair $\{x,y\}$. Various generalizations of the AKLT model to higher spin and multi-dimensional lattices have also been introduced, but we defer this discussion to Section~\ref{sec:gap_methods}.

\subsection{Dynamics} For any finite volume, $\Lambda$, the Heisenberg dynamics $\tau_t^\Lambda(A) = e^{itH_\Lambda}Ae^{-itH_\Lambda}$, $A\in\cA_\Lambda$, gives the time evolution of the system for any $t\in\bR$, and forms a one-parameter group of automorphisms: 
\[
\tau_s^\Lambda\circ\tau_t^\Lambda = \tau_{s+t}^\Lambda, \quad \tau_0^\Lambda ={\rm id} \quad \forall s,t\in\bR.
\]

If the interaction $\Phi$ decays sufficiently fast, which will be discussed below, the dynamics converges strongly to an infinite volume dynamics on $\caA_\Gamma$. For fixed $t\in\bR$, this is shown for $A\in\cA_\Gamma^{\rm loc}$ by proving there exits $\tau_t(A)\in\cA_\Gamma$ such that 
\begin{equation}\label{infinite_dynamics}
  \|\tau_t^{\Lambda_n}(A) -\tau_t(A)\|\to 0 \quad \text{as} \quad \Lambda_n\uparrow \Gamma. 
\end{equation}
This is extended to all quasi-local observables $A\in\cA_\Gamma$ by continuity. The notation $\Lambda_n\uparrow \Gamma$ denotes an increasing and absorbing sequence (IAS), meaning that for all $n$,
\[
\Lambda_n\subseteq \Lambda_{n+1}, \quad \Gamma = \bigcup_{n\geq 1}\Lambda_n\,.
\]
It can be show that the limiting dynamics $\tau_t(A)$ is independent of the IAS.

Similar to the finite-volume case, the infinite-volume dynamics, $\{\tau_t:t\in\bR\}$, forms a strongly-continuous, one-parameter group of automorphisms on $\cA_\Gamma$. Its generator, $i\delta$, is a closed derivation whose dense domain contains $\cA_\Gamma^{\loc}$ as a core. In particular, under the same sufficient decay assumptions on the interaction,
\begin{equation}\label{derivation}
    \delta(A) = \lim_{\Lambda_n\uparrow \Gamma}[H_\Lambda, A]=\sum_{X\in \cP_\Gamma} [\Phi(X), A], \quad A\in\cA_\Gamma^\loc,
\end{equation}
where the last expression above is absolutely norm summable, and $[A,B]=AB-BA$ is the commutator. Formally, one writes $\tau_t=e^{it\delta}.$

When $\Gamma = \bZ^D$, a sufficient condition for establishing \eqref{infinite_dynamics} is the existence of constants $a>0$ and $\theta\in(0,1],$ such that
\begin{equation}\label{interaction-decay}
    \|\Phi\|_{a,\theta} =\sup_{x\in\bZ^D} \sum_{\substack{X\in\cP_{\bZ^D}\\x\in X}}e^{a\cdot\diam(X)^\theta}\|\Phi(X)\|<\infty\,.
\end{equation}
This implies that the interaction terms decay at least as fast as a stretched exponential: $\|\Phi(X)\| \leq \|\Phi\|_{a,\theta}e^{-a\cdot \diam(X)}$. In particular, \eqref{interaction-decay} holds for any translation-invariant, finite-range interactions. Interactions with high enough power-law decay 
\[
\|\Phi(X)\| \leq \frac{C}{(1+\diam(X))^p}, \quad p \geq p(D)
\]
are also sufficient. These criterion can also be generalized to other lattices and metrics.

While not discussed above, one can also consider parameter-dependent interactions, such as time-dependent interactions $\Phi_t(X)^*=\Phi_t(X)\in\cA_X$, $X\in\cP_\Gamma$, $t\in\bR$. It is common that the interaction is continuous in the parameter. The finite-volume Heisenberg dynamics is then defined as family of automorphism generated by the unitary solution a certain differential equation, and sufficient interaction decay once again implies the existence of the limiting dynamics. This will be discussed more in Section~\ref{sec:gap_results}.

The objects used for classifying quantum spin model as being gapped or gapless are the quasi-local algebra, $\cA_\Gamma$, the derivation, $\delta,$ and a state, $\omega:\caA_\Gamma\to\bC$, which we now discuss.

\subsection{States} A state of a $C^*$-algebra $\cA$ is any linear functional $\omega:\cA\to\bC$ that is positive and normalized:
\[
\omega(A^*A) \geq 0, \quad \omega(\idty) = 1.
\]
We first discuss states of finite volume quantum spin systems, and then turn again to the infinite volume setting.

Let $\Lambda\in\cP_\Gamma$ be fixed. For any state $\omega_\Lambda:\cA_\Lambda\to\bC$ there exists a density matrix (which is a non-negative matrix $\rho_\Lambda\in\cA_\Lambda$ with $\tr(\rho_\Lambda)=1$) such that
\[
\omega_\Lambda(A) = \tr(\rho_\Lambda A),\quad \forall \, A\in\cA_\Lambda.
\]
The state is called pure if $\ran(\rho_\Lambda) = \Span\{\psi\}$ for some normalized $\psi\in \cH_\Lambda$. In this case, $\omega_\Lambda(A)=\braket{\psi}{A\psi}$ for all observables, which explains why $\psi\in\cH_\Lambda$ are also referred to as states.

The low-lying energy states associated with an interaction $\Phi$ are of particular interest for categorizing the model into its quantum phase of matter. Since $\dim(\cH_\Lambda)$ is finite, the spectrum of the Hamiltonian $H_\Lambda=H_\Lambda^*$, which set of possible energies, is discrete, real, and given by its set of eigenvalues. Denoting by
\begin{equation}\label{distinct_eigenvalues}
E_\Lambda^0 < E_\Lambda^1< E_\Lambda^2 <\ldots
\end{equation}
the distinct eigenvalues of $H_\Lambda$, the the ground state space is the eigenspace $\cG_\Lambda = \ker(H_\Lambda - E_\Lambda^0)$ associated with the ground state energy, $E_\Lambda^0,$ and a state $\omega_\Lambda(A)=\tr(\rho_\Lambda A)$ is a ground state if $\ran(\rho_\Lambda)\subseteq \cG_\Lambda.$ 

One can show that $\omega_\Lambda$ is a ground state if and only if
\[
\omega_\Lambda(A^*[H_\Lambda,A]) \geq 0, \quad \forall A\in \cA_\Lambda.
\]
Since $i\delta_\Lambda(A) = i[H_\Lambda,A]$ is the generator for the finite-volume Heisenberg dynamics, $\tau_t^\Lambda$, the analogous statement is used to defined ground states of the infinite volume system as there is not a priori an infinite volume Hamiltonian. Namely, a state $\omega:\caA_\Gamma\to\bC$ is a ground state of the of the infinite system dynamics $\tau_t=e^{it\delta}$ if
\begin{equation}\label{ground_state}
  \omega(A^*\delta(A)) \geq 0, \quad \forall A\in\cA_\Gamma^{\rm loc}.  
\end{equation}

The Banach-Alaoglu Theorem guarantees that any collection of finite-volume states $\{\omega_{\Lambda_n}: \Lambda_n\uparrow\Gamma\}$ has weak-$*$ limits which are also linear functionals $\omega:\caA_\Gamma\to\bC$. These limit points inherit the normalization and positive properties of the finite-volume counterparts, and so they are states of the $C^*$-algebra of quasi-local observables. For brevity, these states will be referred to as \emph{limiting states}.  Moreover, under decay assumptions so that \eqref{infinite_dynamics}-\eqref{derivation} hold, if $\omega_{\Lambda_n}$ is a finite-volume ground state for every $n$, the resulting set of weak-$*$ limits will be ground states of the infinite volume dynamics, $\tau_t$, in the sense of \eqref{ground_state}. Such states will be referred to as \emph{limiting ground states}.

The interaction defining the derivation \eqref{derivation} is not unique. For example, adding boundary conditions can produce the same generator for the infinite system dynamics. However, even if the Hamiltonians with different boundary conditions produce the same generator, the limiting ground states for one set of boundary conditions may not be the same for another. In the case of the XXZ model, the addition of kink/anti-kink boundary conditions produces additional limiting ground states than the original interaction. Bratteli, Kishimoto and Robinson provided a classification for all of the infinite volume ground states which can be interpreted in terms of arbitrary boundary conditions imposed on finite volume Hamiltonians. However, this criterion is difficult to work with directly. It suffices to say that one should be aware that, for a fixed set of boundary conditions, it is not guaranteed that the set of limiting ground states is the complete set of the ground states for the generator of the infinite-volume dynamics. Regardless, the limiting ground states are the ones whose spectral gaps can be rigorously studied.

\subsection{Gapped ground states}
That a quantum spin model belongs to a gapped ground state phase is a property of the infinite system. While an algebra of observables, $\cA_\Gamma$, and Heisenberg dynamics, $\tau_t$, can be defined for the infinite system from their finite volume counterparts, it is not possible to similarly construct an infinite volume Hamiltonian or Hilbert space as the corresponding finite volume objects do not converge in the thermodynamic limit. However, they can be obtained from states of the quasi-local algebra. 

The Gelfand–Naimark–Segal (GNS) construction guarantees that for any state $\omega:\caA_\Gamma\to\bC$, there exists a Hilbert space $\cH$, cyclic vector $\Omega\in\cH$, and $*$-representation, $\pi:\cA_\Gamma \to \cB(\cH)$, onto the bounded linear operators of $\cH$ so that
\begin{equation}\label{GNS}
    \omega(A) = \braket{\Omega}{\pi(A)\Omega},\quad \forall A\in \cA_\Gamma.
\end{equation}
The subspace $\pi(\cA_\Gamma)\Omega$ is dense in $\cH,$ and the GNS triple $(\cH,\pi,\Omega)$ is unique up to unitary equivalence. 

When $\omega$ is invariant under the dynamics, i.e. $\omega\circ \tau_t = \omega$ for all $t\in \bR$, the uniqueness implies the existence of unitaries $\{U_t : t\in \bR\}$ such that for each $t$, $ U_t\Omega = \Omega$ and 
\[
\pi(\tau_t(A)) = U_t^*\pi(A)U_t, \quad A\in\cA_\Gamma\,.
\]
These form a strongly continuous, one parameter group of unitaries, which is generated by a self-adjoint operator $H$ by Stone's Theorem, i.e. $U_t=e^{-itH}$. This operator, called the GNS Hamiltonian, is unbounded and its domain contains $\pi(\cA_{\Gamma}^{\rm loc})\Omega$ as a dense core. Moreover, $\Omega\in\ker(H)$ since $U_t\Omega=\Omega$ implies
\[
-iH\Omega = \lim_{t\to 0}\frac{U_t\Omega -\Omega}{t} = 0.
\]

This applies to ground states as they are invariant under the dynamics, and \eqref{ground_state} can be used to show that $H\geq 0$. Therefore, $\min\spec(H) = 0$ and the \emph{spectral gap above the ground state} is
\[
\gap(H) = \sup\{\delta>0 : \spec(H) \cap (0,\delta) = \emptyset\},
\]
where one sets $\gap(H)=0$ if the set above is empty. We say that a ground state is a \emph{gapped ground state} if its GNS Hamiltonian has a positive gap, $\gap(H)>0.$

\section{Uniformly gapped quantum spin models}\label{sec:gap_methods}

Analyzing the gap of the GNS Hamiltonian directly is challenging since an explicit expression for the operator is not commonly known. However, a uniform lower bound on the spectral gaps associated to an IAS of finite volume Hamiltonians can be lifted to a lower bound of the GNS Hamiltonian of any limiting ground state.

The spectral gap above the ground state of a finite-volume Hamiltonian is the difference between its first excited state and ground state energies:
\begin{equation}\label{finite_gap}
    \gap(H_\Lambda) = E_\Lambda^1-E_\Lambda^0, \quad \Lambda\in\cP_\Gamma,
\end{equation}
where we use the notation from \eqref{distinct_eigenvalues}. It is clear that $\gap(H_\Lambda)>0$ for any $\Lambda\in\cP_\Gamma$. However, as one is interested the behavior of the spectrum in the thermodynamic limit, the notion of a uniform gap is concerned with where this has a positive uniform lower bound over an IAS. Concretely, a model is \emph{uniformly gapped} if there exists $\Lambda_n\uparrow \Gamma$ so that
\begin{equation}\label{uniform_gap}
\inf_n \gap(H_{\Lambda_n}) >0. 
\end{equation}

If $\omega$ is a limiting ground state of a sequence $\Lambda_n\uparrow\Gamma$, and the interaction $\Phi$ has sufficient decay so that \eqref{infinite_dynamics}-\eqref{derivation} hold, then
\begin{equation}\label{uniform_bound}
    \gap(H)\geq \limsup_n\gap(H_{\Lambda_n})
\end{equation}
where $H$ is the GNS Hamiltonian of $\omega$. Hence, a positive lower bound on the uniform gap implies $\omega$ is a gapped ground state. 

One can also consider whether or not the local Hamiltonians with boundary conditions have a uniform gap. Under certain relatively mild constraints of the boundary conditions, the analogous version of \eqref{uniform_bound} holds, and the uniform gap will again produce a lower bound of the GNS Hamiltonian of any limiting ground state of the model with boundary conditions. For example, by considering the XXZ model with kink/anti-kink boundary conditions, Koma and Nachtergaele were able to prove uniform gaps for a much wider class of limiting ground states than those of the original interaction.  Moreover, for periodic boundary conditions, a uniform gap will produce a lower bound on the so-called bulk gap of the system. This is particularly useful for topological insulators, whose uniform gap will always vanish for open boundary conditions, but is expected to be strictly positive for periodic boundary conditions.

All general methods for (directly) proving a lower bound on $\gap(H_\Lambda)$ can be traced back to the AKLT model. The main difference between this model and the spin-$1$ antiferromagnetic Heisenberg model, which AKLT leveraged to rigorously prove a uniform gap, is that the interaction is frustration free. The techniques for proving lower bounds on spectral gaps discussed in this section only hold for frustration free models. Stability results, which can produce lower bounds on the spectral gap of non-frustration free models, will be discussed in Section~\ref{sec:Stability}.

An interaction $\Phi$ is called \emph{frustration free} if every ground state $\psi\in \cG_\Lambda$, $\Lambda\in\cP_\Gamma,$ simultaneously minimizes the energy of all its associated interaction terms:
\[
\Phi(X)\psi = \mu_X\psi, \quad \mu_X = \min\spec(\Phi(X)) \quad \forall X\subseteq \Lambda.
\]
Said differently, for all $\Lambda\in\cP_\Gamma,$
\[
\cG_\Lambda = \bigcap_{X\subseteq \Lambda} \ker(\Phi(X)-\mu_X), \quad \forall \Lambda\in\cP_\Gamma.
\]


For a frustration free interaction, $\Phi$, an infinite-volume state $\omega:\caA_\Gamma \to \bC$ is called a ``frustration free state'' if
\begin{equation}\label{ff_state}
    \omega(\Phi(X)) = 0 \quad \forall X\in\cP_\Gamma.
\end{equation}
For frustration free interactions, an infinite-volume state $\omega$ is a limiting ground state if and only if $\omega$ is a frustration free state.

\subsection{Methods for establishing uniform gaps}

The objective for establishing a positive uniform gap is to produce a lower bound on $\gap(H_{\Lambda})$ that is independent of the system size. For finite-range, translation-invariant interactions, the intuition that low-lying, bulk excitations should not involve sites that are too far apart motivates a general approach for proving such a lower bound. Namely, if the size of a generic region where an excitation is localized can be effectively bounded in space, then the spectral gap of Hamiltonian of such a region can be used to produce a lower bound the spectral gap of the Hamiltonian defined on any sufficiently larger region. For frustration free models, there are two main classes of methods for proving spectral gaps. For clarity, we explain their main ideas for models defined on $\Gamma=\bZ.$

The first class of spectral gap methods uses ground state projections to effectively localized excitations. Let $\Lambda=[a,b]\subseteq \bZ$ and denote by $G_{\Lambda'}$ the orthogonal projection onto $\cG_{\Lambda'}\otimes\cH_{\Lambda\setminus \Lambda'}$ for any sub-interval $\Lambda'\subseteq \Lambda$. Partition $\Lambda=\Lambda_L\cup\Lambda_M\cup\Lambda_R$ into a left, middle, and right interval, and define sub-intervals $\Lambda'=\Lambda_M\cup\Lambda_R$ and $\Lambda''=\Lambda_L\cup\Lambda_M$. The main objective with these methods is to produce a bound of the form
\begin{equation}\label{MM_cond}
\|G_{\Lambda'}(\idty - G_\Lambda)G_{\Lambda''}\| <  \epsilon
\end{equation}
where $0<\epsilon<1$ is the threshold indicating if an excitation has been effectively localized. Indeed, one interpretation of  \eqref{MM_cond} is that the excitation of any $\psi \in \cG_\Lambda^\perp\cap (\cG_{\Lambda''}\otimes\cH_{\Lambda\setminus\Lambda''})$ is localized to $\Lambda'$ up to error $\epsilon.$ The precise value of $\epsilon$ that is sufficient to prove a positive lower bound on the spectral gap is explicit, and depends on how the particular gap method is applied. 

Lower bounds on the uniform gap associated with an IAS $\Lambda_n\uparrow \Gamma$ will be produced if \eqref{MM_cond} can be proved for enough suitably chosen volumes $\Lambda$ and subvolumes $\Lambda',\Lambda''$. An important criterion is that the collection of regions used to localize excitations support every nonzero interaction term, i.e. for all $X\subseteq \Lambda_n$ for which $\Phi(X)\neq 0$, there exists $\Lambda'\subseteq\Lambda_n$ such that $X\subseteq \Lambda'$. The lower bound on the uniform gap will depend on the spectral gaps of the localization regions, $\gap(H_{\Lambda'})$, so these should be chosen independent of $\Lambda_n.$

Specific techniques that use ground state projections to localize excitations include the martingale method \citep{Nachtergaele1996} as well as its many variations, the Fannes-Nachtergaele-Werner (FNW) approach \citep{Fannes1992}, and local gap bounds such as the detectability lemma. These methods can also be applied to any dimensional lattice and to Hamiltonians with various boundary conditions.

The second class of gap methods is finite size criteria, and one of the most celebrated is \citep{Knabe1988}. The models considered in that work were spin chains defined in terms of translation-invariant, nearest-neighbor, frustration free interactions whose terms were orthogonal projections. For such models, it was proved that, for any $L>2n\geq 4$, the spectral gap of the Hamiltonian $H_{[1,L]}^{\rm per}$ with periodic boundary conditions can be bounded below by
\begin{equation}\label{FSC}
    \gap(H_{[1,L]}^{\rm per}) \geq \frac{n}{n-1}\left(\gap(H_{[1,n+1]})-\frac{1}{n}\right).
\end{equation}

More generally, finite size criteria establish that a model is uniformly gapped when the spectral gap of a finite-volume Hamiltonian surpasses a certain threshold, e.g. $\gap(H_{[1,n+1]})>1/n$. Other finite size criteria in the spirit of Knabe have successfully (1) reduced the necessary gap threshold (2) extended their applicability to multi-dimensional lattices (3) produced lower bounds on the spectral gap of Hamiltonians with more general boundary conditions. Uniform gaps for more general finite range interactions can also be analyzed after coarse graining the interaction.

Common to all of the methods discussed above is that the lower bound produced on the spectral gap of the Hamiltonian on $\Lambda$ depends on the spectral gap of the Hamiltonian on a subsystem, $\Lambda'\subseteq \Lambda$. Regardless of the choice of boundary conditions for the Hamiltonian on $\Lambda$, the Hamiltonian on $\Lambda'$ naturally has open boundary conditions (OBC). This makes it challenging to proving uniform gaps for models that have low-lying edge excitations. Edge states are excited states of the Hamiltonian where the excitation is localized to the boundary of the finite volume system. For example, topological insulators have low-lying edge modes that will produce a vanishing uniform gap for the model with OBC, but nevertheless have a positive bulk gap. In this and other similar cases, gap estimates like \eqref{FSC} may not accurately reflect the behavior or presence of the bulk gap. 

Under certain conditions, the gap methods discussed above can also be applied to Hamiltonians that are restricted to subspaces. One path to circumvent edge modes is to separate the ground states of $H_\Lambda^{\rm per}$ and the edge states of $H_{\Lambda'}\otimes \idty_{\Lambda\setminus\Lambda'}$, $\Lambda'\subseteq \Lambda$, into different invariant subspaces of $H_\Lambda^{\rm per}$. Then, one can  apply spectral gap methods to the subspace containing the ground states, and techniques for bounding ground state energies that do not require considering subsystems $\Lambda'\subseteq \Lambda$ to the subspace with the edge states. Combining these two bounds results in a lower bound on the bulk gap, $\gap(H_\Lambda^{\rm per})$, independent of the edge modes of the model with OBC.

Successfully applying the above methods proves that the frustration free states $\omega:\caA_\Gamma\to\bC$ of a frustration free interaction are gapped ground states. The majority of models for which uniform gaps have been proved are on the one-dimensional lattice, $\bZ$. However, rigorous results for models on multi-dimensional lattices have appeared more often in recent years, and typically result from iterations or novel combinations of gap methods previously developed for $\bZ$. It is worth noting that models whose interaction terms are comprised of orthogonal projection that pairwise commute, such as Kitaev's quantum double models and the Levin-Wen models, are necessarily uniformly gapped since it is easy to show that $\gap(H_\Lambda) \geq 1$ for any $\Lambda\in\cP_\Gamma$. The Hamiltonians of these models are called commuting local projection (CLP) Hamiltonians and are of interest for topological quantum computing.

\subsection{Results on decidability and likelihood of a gap}

In this section, we discuss the following two questions:
\begin{enumerate}
    \item  Is it possible to decide algorithmically whether or not a quantum spin model belongs to a gapped phase?
    \item  How likely is it for a generic quantum spin model to be gapped?
\end{enumerate}

The decidability of the spectral gap was answered for quantum spin models on $\Gamma =\bZ^D$, $D\geq 2$, with translation invariant, nearest neighbor, frustration free interactions in \citep{Cubitt2015}. Specifically, a quantum spin interaction was constructed for which the problem of determining if the model is gapped could be translated into a halting problem of a Turing machine. Since the latter is a known to be algorithmically undecidable, so too is the spectral gap question for the constructed interaction. Hence, no single algorithm can determine for all interactions whether they belong to a gapped ground state phase. A similar result also holds for $D=1.$

A related but different question is how likely it is for a generic quantum spin model to be gapped (in a probablistic sense). Under weak assumptions on the distribution of the interaction terms, it has been shown that a nearest neighbor model defined by a random interaction would be almost surely gapless. However, this excludes translation invariant models, for which the likelihood of a gap is less stark.

A gap is more typical for translation invariant, frustration free interactions. If the rank $r$ of the interaction terms is not too large, then the probability that the model is uniformly gapped is strictly positive. The upper-bound on $r$ only depends on the spatial dimension of the lattice, and the dimension of the onsite Hilbert space $n=\dim(\cH_{\{x\}})$. For $\Gamma = \bZ^D$, an explicit bound is
\[
r\leq \min \left\{\frac{\lfloor\frac{n^2}{4}\rfloor}{D}, \; \frac{e^{-1}}{4D-1}n^2\right\}.
\]
An interesting note is that the example used to show the gap is undecidable has an interaction with large rank compared to the generic gap result above. See \citep{Jauslin:2022} and references therein for more information about the generic behavior of the gap.

\subsection{Progress on long-standing conjectures}

Several generalizations of the AKLT model were also introduced in \citep{Affleck1988}. One of the most commonly considered is inspired from scaling and shifting the one-dimensional AKLT interaction from \eqref{aklt} to find
\[
\frac{1}{2}\Phi^{AKLT}\left(\{x,x+1\}\right) + \frac{1}{3} = P_{\{x,x+1\}}^{(2)}
\]
where $P^{(2)}_{\{x,x+1\}}$ is the orthogonal projection onto the spin-2 subspace, i.e., subspace of maximal total spin, associated with the tensor product of the two spin-1 sites $x$ and $x+1.$ This lead to the valence-bond-state (VBS) representation of the ground state space that was used to prove a positive uniform gap for the spin-1 AKLT chain. This generalization extends the VBS ground state representation to higher dimensional lattices.

Given a lattice where $\deg(x) = d$ at each site $x\in\Gamma$, a particle of spin $s=d/2$ is fixed at each site, and the nearest-neighbor AKLT interaction on $\Gamma$ is defined by
\[
\Phi^{AKLT}\left(\{x,y\}\right) = P_{\{x,y\}}^{(d)}, \quad |x-y|=1.
\]
where $P_{\{x,y\}}^{(d)}$ is the orthogonal projection onto the subspace of spin-$d$ between the two spin-$d/2$ particles at sites $x$ and $y.$ Affleck, Kennedy, Lieb, and Tasaki conjectured that the model would be gapped if the degree of the lattice was sufficiently small, and otherwise would exhibit N\'eel order (which, by Goldstone's Theorem, implies the model is gapless). Specifically, it has been conjectured that the AKLT model on the two-dimensional honeycomb lattice and square lattice are gapped, which is supported by numerical evidence. Proving this conjecture is significant since the ground states of AKLT models have a Tensor Network State (TNS) representation, and this may provide a pathway for analyzing the gap of other models on multi-dimensional lattices with TNS ground states. We point the interested reader to \citep{Cirac2021} for a review of tensor network states.

Significant progress on the spectral gap conjecture was only accomplished thirty years later, culminating in two works which combined numerical and analytic techniques to establish a uniform gap for the honeycomb model. The first, \citep{Lemm2020}, combined a finite size criterion with a density matrix renormalization group (DMRG) algorithm. DMRG is designed to be well suited for models with TNS ground states. The second, \citep{Pomata2020}, used the TNS representation of the ground state space to apply the FNW approach coupled with a Lanczos numerical method.

Another interesting class of models where progress on a spectral gap conjecture has been made is Haldane's pseudopotentials \citep{Haldane1983}, which  are Hamiltonian models for the fractional quantum Hall effect (FQHE). These models are labeled by their ground state filling factor $\nu$, and the two-dimensional manifold on which they are defined. They are designed to have a Laughlin many-body wave function as a maximally filled ground state, and it has been conjectured they share other features expected of the FQH phase. This includes the incompressibility of the FQH fluid, and a ground state gap above which are fractional excitations (i.e. anyons). While numerical evidence supports the spectral gap conjecture for small fillings $\nu$, a rigorous proof is still needed.

In second quantization, Haldane pseudopotentials take the form of a one-dimensional lattice model. The interaction is not finite range, but does satisfy \eqref{interaction-decay}. A uniform gap for a finite range truncation of the interaction that should well-approximate the $\nu=1/3$ cylinder model has been proved. The first such proof produced a uniform bulk gap that converged to zero as the cylinder radius $R\to 0$ despite numerics indicating the bulk gap remained open in this limit. This suggested the lower bound was reflecting the presences of edge modes of the model with open boundary conditions. To circumvent these edge modes, the invariant subspace strategy discussed in Section~\ref{sec:gap_methods} was introduced and successfully applied to produce a uniform bulk gap that remained open as $R\to 0$ \citep{Warzel2023}.

\section{Gapped ground state phases of short range interactions} \label{sec:gap_results}

In this section, we review properties that hold for gapped quantum spin models, including: exponential decay of correlations, area laws, and adiabatic theorems. The common thread connecting these results is that they all assume (1) a uniform gap above the ground state energy for the local Hamiltonians, and (2) locality estimates on the Heisenberg dynamics. Lieb-Robinson bounds produce explicit bounds on this locality, and have proved to be an invaluable tool for analyzing a wide variety of questions related to quantum lattice models, including proofs of all of the above listed results. 

\subsection{Lieb-Robinson bounds}
A consequence of the non-relativistic nature of quantum many-body theory is that, with the exception of some trivial cases, the evolution of a local observable $A\in\cA_\Gamma^{\loc}$ under the Heisenberg dynamics is no longer local, i.e. $\tau_t(A)\notin\cA_\Gamma^{\loc}$ for any $t\neq 0$. However, for interactions with sufficient decay, $\tau_t(A)$ is well-approximated by a local observable whose support grows like a light cone governed by time and the rate of decay of the interaction. 

The quasi-locality of time-evolved observable is a consequence of a Lieb-Robinson bound \citep{Lieb1972}, which produces an upper bound on the operator norm of the commutator $[\tau_t(A), B]$ for any pair of spatially separated local observables $A\in\cA_X$ and $B\in\cA_Y$, $X\cap Y=\emptyset$. As an example, for an interaction satisfying \eqref{interaction-decay} and any $0<a'<a$, there exists positive constants $C_i:=C_i(a-a')$, $i=1,2$, independent of $X$ and $Y$, so that
\begin{equation}\label{LR_bound}
\|[\tau_t(A),B]\|\leq C_1|X|\|A\|\|B\|e^{C_2\|\Phi\|_{a',\theta}|t|-a'd(X,Y)^\theta}\
\end{equation}
for all $A\in\cA_X$ and $B\in\cA_Y.$ In the case $\theta=1$, the constant $v_{a'}=C_2\|\Phi\|_{a',1}/a'$ is called the ``Lieb-Robinson velocity.'' Lieb-Robinson bounds also hold, e.g., for time-dependent interactions, interactions with power-law decay, lattice boson models, and open quantum systems. We point the reader to \citep{Nachtergaele2019} and references therein for more information on this topic.

One way to obtain a local approximation of $\tau_t(A)$ is through the normalized partial trace. For any $\Lambda\in\cP_\Gamma$, this is the map $\Pi_\Lambda:\caA_\Gamma \to \cA_\Lambda$ which is defined on local observables by
\[
\Pi_\Lambda = \idty_{\Lambda}\otimes\bigotimes_{x\in\Gamma\setminus\Lambda}\rho_x,\quad \rho_x(A) = \frac{\Tr(A)}{\dim(\cH_x)}\;\;\forall A\in\cA_{\{x\}}.
\]
The error of the local approximation $\Pi_\Lambda(A)$ for any $A\in\cA_\Gamma$ is bounded in terms of a commutator estimate:
\begin{equation} \label{Local_approximation}
    \|A-\Pi_\Lambda(A)\| \leq \sup_{\substack{0\neq B\in\cA_\Gamma^{\rm loc} \\ \supp(B)\subseteq \Gamma\setminus\Lambda}}\frac{\|[A,B]\|}{\|B\|} \leq 2\|A-\Pi_\Lambda(A)\|\,.
\end{equation}
 Hence, given any $\epsilon>0$, $t\in\bR$ and $A\in\cA_X$, $X\in\cP_\Gamma$,
\[\|\tau_t(A) - \Pi_{X_t(\epsilon)}(\tau_t(A))\|<\epsilon\|A\|\]
where $X_t(\epsilon)\in \cP_\Gamma$ is the smallest volume so that right hand side of \eqref{LR_bound} is smaller than $\epsilon\|A\|\|B\|$ for all $Y\subseteq \Gamma\setminus X_t(\epsilon).$ 

Other product states defined in terms of density matrices with full rank can also be used to define localizing maps.

\subsection{Decay of correlations and area laws}

Correlations bounds measure how independent measurements in two spatially separated parts of the system are from one another. The expectation that ground states of gapped quantum spin models should exhibit exponential decay of correlations was long expected and motivated by a similar result in relativistic quantum field theory for the mass gap. The commutativity of spatially separated observables was harnessed to prove exponential decay of correlations in the case of the bulk gap. For quantum spin systems, it is the quasi-locality of the Lieb-Robinson bound. This result holds for both finite and infinite quantum spin models \citep{Hastings2006,Nachtergaele2006}. As the GNS representation equally applies to finite systems, we state the result in this context.

Suppose that $\omega$ is a ground state of the dynamics generated by an interaction satisfying $\|\Phi\|_{a,1}<\infty$ for some $a>0$ whose GNS Hamiltonian satisfies $\gamma=\gap(H)>0$. Let $P_0$ denote the orthogonal projection onto the ground state space, $\ker(H)$. Then, for any local observables $A\in\cA_X$ and $B\in\cA_Y$ such that $P_0\pi(B)\Omega = P_0\pi(B^*)\Omega=0$
\begin{equation}\label{correlations}
    |\omega(AB)-\omega(A)\omega(B)|\leq C(X,Y,\gamma)\|A\|\|B\|e^{-\mu d(X,Y)},
\end{equation}
where, for a constant $C_a$ only depending on $a$,
\[\mu = \frac{a\gamma}{4\|\Phi\|_{a,1}C_a+\gamma}.\]

The assumptions imply that $\omega(B) =0$ and so the left hand side of \eqref{correlations} can be simplified. Moreover, at a penalty to the exponential decay, $\mu$, the bound in \eqref{correlations} can be generalized to a bound of $|f(ib)|=|\omega(A\tau_{ib}(B))|$ for any $0\leq b\gamma \leq 2\mu d(X,Y).$ The proof of this is based on using analyticity of $f(z)$ in the upper-half plane to rewrite $f(ib)$ in terms of Cauchy's integral formula. The positive spectral gap implies that a portion of this contour integral vanishes, yielding
\[|f(ib)|\leq \limsup_{T\to\infty}\left|\frac{1}{2\pi i}\int_{-T}^T\frac{f(t)}{t-ib}dt\right|\,.\]
After rewriting $f(t)$ using the commutator $[A,\tau_t(B)]$, one can use the Lieb-Robinson estimates to bound the remaining term. A similar application of Lieb-Robinson bounds was also used to proved a multi-dimensional version of the LSM Theorem \citep{Hastings2004}. 

There is a long standing belief that states with exponential decay of correlations should satisfy an area law for the entanglement entropy. Roughly speaking, an area law holds for a state $\rho_\Lambda$ if for any convex region $\Lambda'\subseteq \Lambda$, the von Neumann entropy grows at most like the boundary of $\Lambda'$ rather than the volume,
\[
\cS(\rho_{\Lambda'}) = -\Tr(\rho_{\Lambda'}\log(\rho_{\Lambda'})) \propto \mathcal{O}(|\partial \Lambda'|),
\]
where $\rho_{\Lambda'}$ is the reduced density matrix of $\rho_\Lambda$ onto $\Lambda'$. The first such proof used Lieb-Robinson bounds and the exponential decay of correlations to establish an area law for ground states of gapped one-dimensional lattice systems \citep{Hastings2007}. However, this bound does not scale to higher dimensional systems as it depends exponentially on $n=\dim(\cH_x)$.

Several years later, an alternate approach proved an area law with an exponential improvement for one-dimensional frustration free, gapped ground states of nearest-neighbor Hamiltonians; namely,
\[
\cS(\rho_{\Lambda'}) \leq \mathcal{O}(1) \mu^3\log^8(\mu),\quad \mu = J\log(n)/\gamma  
\]
where $J = \sup_{x}\|\Phi(\{x,x+1\})\|$. This proof uses ground state approximation techniques similar to the those used in the dectability lemma as well as combinatorial arguments \citep{Arad:2012}, and has paved the way for other area laws, including one for two-dimensional systems \citep{Anshu2022}.

\subsection{Gapped phases and the quasi-adiabatic continuation} \label{sec:QAC} 
Two gapped quantum spin interactions $\Phi_i$, $i=0,1$, are in the same quantum phase if they can be connected along a uniformly gapped, (piecewise) smooth path of interactions $\Phi_s$, $0\leq s\leq 1$, \citep{Chen2010,Chen2011}. Here, $\Phi_s$ is smooth if each interaction term is differentiable and pointwise continuous, meaing that for all $X\in\cP_\Gamma$
\[\|\Phi_s'(X)-\Phi_{s_0}'(X)\|\to 0 \quad \text{as}\quad s\to s_0\]
where $\Phi_s'(X) = \frac{d}{ds}\Phi_s(X).$  The notion of a uniformly gapped path of interactions is slightly more general than the one from Section~\ref{sec:gap_methods}. This allows for a negligible amount of energy splitting in the ground state for the local Hamiltonians that does not effect the ground state in the thermodynamic limit. Letting $H_\Lambda(s) = \sum_{X\subseteq\Lambda}\Phi_s(X)$ and $E_\Lambda^0(s) = \min\spec(H_\Lambda(s))$ denote the local Hamiltonian and its ground state energy, respectively, a path $\Phi_{s}$, $0\leq s \leq 1$, is uniformly gapped if there exists $\gamma>0$, an IAS $\Lambda_n\uparrow\Gamma$, and non-negative sequence $\epsilon_n\to 0$ so that for all $n\geq 1$,
\begin{equation}\label{UG_spec}
    \spec(H_{\Lambda_n}(s)) \subseteq [E_{\Lambda_n}^0(s),E_{\Lambda_n}^0(s)+\epsilon_n]\cup [E_{\Lambda_n}^0(s)+\epsilon_n+\gamma,\infty).
\end{equation}

One invaluable tool used in this classification, as well as many subsequent works on the classification of quantum phases, is the quasi-adiabatic continuation introduced in \citep{Hastings2005}. Under relatively mild assumptions, this family of automorphisms $\alpha_s:\caA_\Gamma\to\caA_\Gamma$, $s\in[0,1]$, transforms the ground states of $\Phi_0$ into ground states of $\Phi_s$, and can be used to help characterized properties that are common to all states along the path.

 The construction of the quasi-adiabatic continuation follows similarly that of the infinite dynamics; one first considers finite systems and then shows that they converge in the thermodynamic limit \citep{Bachmann2012}. A decay assumption is also needed on the interaction as well as its derivative. A sufficient condition is that there exists $a>0$ and $\theta\in(0,1]$ for which
\begin{equation}\label{TD_decay}
    |||\Phi|||_{a,\theta} =\sup_{0\leq s \leq 1}\left( \|\Phi_s\|_{a,\theta} +  \|d\Phi_s\|_{a,\theta}\right)<\infty,
\end{equation}
where $d\Phi_s(X) = |X|\Phi_s'(X)$ for all $X\in\cP_\Gamma,$ and $\|\cdot\|_{a,\theta}$ is as in \eqref{interaction-decay}. 
Under this assumption, the $s$-dependent generator of the quasi-adiabatic continuation for any $\Lambda\in\cP_\Gamma$, $D_\Lambda(s)$, is then defined by the weighted integral
\begin{equation}\label{hastings_gen}
    D_\Lambda(s) = \int_{-\infty}^{\infty}W_\gamma(t) e^{itH_\Lambda(s)}H_\Lambda'(s)e^{-itH_\Lambda(s)}dt\,.
\end{equation}
The weight function $W_\gamma\in L^1(\bR)$ is taken so that its distributional derivative has the form $W_\gamma'(t) = w_\gamma(t) + \delta_0(t)$, where $w_\gamma(t)$ decays like a stretched exponential and whose Fourier transform satisfies $\supp(\hat{w}_\gamma)\subseteq [-\gamma,\gamma]$. For an explicit definition of these functions and some of their properties, see \citep{Bachmann2012, Nachtergaele2019}.

The operator $D_\Lambda(s)$ is pointwise self-adjoint and continuous for all $s\in[0,1]$. Interpreting $s$ as a time-parameter, the quasi-adiabatic continuation is the Heisenberg dynamics
\[
\alpha_s^\Lambda(A) = U_\Lambda(s)^* A U_\Lambda(s), \quad 0\leq s \leq 1
\]
that solves the differential equation
\[
  \frac{d}{ds}U_\Lambda(s) = -iD_\Lambda(s)U_\Lambda(s),\quad U_\Lambda(0) = \idty_\Lambda.  
\]

The inspiration for this choice of $D_\Lambda(s)$ comes from a desire to construct an automorphism, $\alpha_s^\Lambda$, that (1) satisfies a Lieb-Robinson bound, and (2) produces a `flow' of spectral projections $P_\Lambda(s)$, $0\leq s\leq 1$, that are associated with portions of the spectrum that are separated from the remaining spectrum by a distance $\gamma.$ Explicitly, assume $P_\Lambda(s)$ is the spectral projection associated with $\spec(H_\Lambda(s))\cap \Sigma_{\Lambda}^1(s) \neq\emptyset$ where $\Sigma_{\Lambda}^1(s)$ is an interval whose endpoints depend smoothly on $s$, and
\[
\spec(H_\Lambda(s)) \subseteq \Sigma_{\Lambda}^1(s)\cup\Sigma_{\Lambda}^2(s), \quad d(\Sigma_{\Lambda}^1(s)\cup \Sigma_{\Lambda}^2(s)) \geq \gamma\,.
\]
Then, the quasi-adiabatic evolution generated by $D_\Lambda(s)$ satisfies 
\begin{equation}\label{spectral_flow}
    \alpha_s^\Lambda(P_\Lambda(s)) = P_\Lambda(0), \quad 0 \leq s \leq 1.
\end{equation}
Note that when $\Lambda = \Lambda_n$ for some $n$ as in \eqref{UG_spec}, the above holds for the spectral projection onto $\Sigma_{\Lambda_n}^1(s) = [E_{\Lambda_n}(s),E_{\Lambda_n}(s)+\epsilon_n]$.

The assumptions on the weight function, $W_\gamma$, and interaction, $\Phi_s$, are sufficient to guarantee that the partial trace can be used to rewrite $D_\Lambda(s)$ as the local Hamiltonian associated with an $s$-dependent interaction that decays at least as fast as a stretched exponential. As a consequence, a time-dependent version of the theory from Section~\ref{sec:setup} holds, and $\alpha_s^\Lambda$ converges in the thermodynamic limit:
\[
\alpha_s^{\Lambda_n}(A) \to \alpha_s(A), \quad \forall A\in\cA_{\Gamma}^{\loc}.
\]
As in the case of the dynamics, this is independent of the sequence $\Lambda_n\uparrow \Gamma$, and can be extended to all observables $A\in\cA_\Gamma.$ 

The infinite-volume quasi-adiabatic continuation constructs a flow of states $\omega_s = \omega_0\circ\alpha_s$ such that if \eqref{UG_spec} holds, then $\omega_s$ is a pure gapped ground state of the dynamics generated by $\Phi_s$ if and only if $\omega_0$ is a pure gapped ground state of the dynamics generated by $\Phi_0$. A precise statement for limiting ground states is as follows, but this also holds for an arbitrary set of pure gapped ground states of $\Phi_0$.

Fix an increasing and absorbing sequence $\Lambda_n\uparrow \Gamma$ such that the associated Hamiltonians $H_{\Lambda_n}(s)$ is a uniformly gapped path in the sense of \eqref{UG_spec}, and denote by $P_{\Lambda_n}(s)$ the othogonal projection onto the eigenstates of $H_{\Lambda_n}(s)$ whose energies lie in the interval $[E_{\Lambda_n}(s),E_{\Lambda_n}(s)+\epsilon_n]$. Finally, let
$\cG_s^{\Lambda_n}$ be the set of all finite volume states whose range is supported on this interval, 
\[
\cG_s^{\Lambda_n}=\left\{\omega:\caA_{\Lambda_n}\to \bC\,|\, \omega \;\text{ a state s.t.}\; \omega(P_{\Lambda_n}(s)) = 1\right\}.
\]
Correspondingly, let $\delta_s$ denote the closed derivation generating the infinite dynamics of $\Phi_s$ as in \eqref{derivation}, and let $\cG_s$ be the set of all limiting states 
\[
\cG_s = \left\{\omega:\cA_\Gamma\to \bC\;\text{a state}\,|\, \omega_{\Lambda_n}\stackrel{\ast}{\rightharpoonup} \omega,\omega_{\Lambda_n}\in\cG_s^{\Lambda_{n}}\right\}\,.
\]
The condition that $\epsilon_n \to 0$ from \eqref{UG_spec} implies that $\cG_s$ is a set of ground states for $\delta_s,$ and the uniform gap implies that, if the kernel of the GNS Hamiltonian for $\omega_0\in\cG_0$ is one-dimensional, then the gap of the GNS Hamiltonian of $\omega_s=\omega_0\circ\alpha_s$ is at least $\gamma$ for all $s\in[0,1]$. Moreover, the spectral flow automorphism identically maps these sets onto one another:
\[
\cG_s = \cG_0\circ \alpha_s, \quad s\in[0,1] \,.
\]

\subsection{The many-body adiabatic theorem} There are many important applications of the quasi-adiabatic continuation in addition to those that are discussed here. For example, it was central in the first rigorous proof of the quantization of the Hall conductance for interacting electron systems on the torus \citep{Hastings2015}. The analysis of this work was based on using the quasi-adiabatic continuation to construct a bundle of states whose adiabatic curvature is the Hall conductance, an identity that is a consequence of Kubo's formula for linear response. However, Kubo's formula was also not rigorously established at the time. This was resolved with the help of a many-body version of the adiabatic theorem. Adiabatic theorems capture the behavior of slowly driven systems. Namely, when a ground state is slowly evolved under the one-parameter family of unitaries generated by a gapped Hamiltonian, the resulting state after this driving should approximately be a ground state.  We close this section with the adiabatic theorem from \citep{Bachmann2018}.

We consider a similar setting as in Section~\ref{sec:QAC}. Suppose that $\Phi_s$, $0\leq s \leq 1$ is an interaction for a model on $\Gamma = \bZ^D$ that is $D+k+1$-times smoothly differentiable for some $k\geq 1$ and whose derivative is compactly supported on $s\in(0,1)$. Moreover, assume the ground state energy is separated from the rest of the spectrum by a gap, i.e.,
 \[\spec(H_\Lambda(s)) \subseteq \left\{E_\Lambda^0(s)\right\}\cup \left[E_\Lambda^0(s)+\gamma,\infty\right)
 \] 
where $\gamma>0$ is independent of $s$ and $\Lambda.$ Let $P_\Lambda(s)$ denote the orthogonal projection onto the ground state space of $H_\Lambda(s)$, and for any $\epsilon>0$ let $U_\Lambda^\epsilon(s)$ denote the unitary evolution that solves
\[
i\epsilon\frac{d}{ds}U_\Lambda^\epsilon(s) = H_\Lambda(s)U_\Lambda^{\epsilon}(s), \quad U_\Lambda^{\epsilon}(0)=\idty_\Lambda\,.
\]
For any $\psi(0)\in\cH_\Lambda$, the state $\psi^\epsilon(s) = U_\Lambda^\epsilon(s)\psi(0)$ solves the analogous Schr\"odinger equation. 

Even if the initial state, $\psi(0)$, is a ground state, it is not generally true that the evolved state is a ground state, i.e. $\psi^\epsilon(s)\notin \ran (P_\Lambda(s))$. However, if the driving is sufficiently slow, which is equivalent to taking $\epsilon$ small, the expectations of observables in the state $\psi^\epsilon(s)$ will be almost indistinguishable from those in a ground state. Concretely, given $\psi(0)\in \ran (P_\Lambda(0))$ and $\epsilon>0$, there exists a state $\psi(s)\in\ran (P_\Lambda(s))$ and constant $C$ independent of $\Lambda$ and $\epsilon$ so that for all observables $A\in\cA_\Lambda$,
\[
|\braket{\psi^\epsilon(s)}{A\psi^\epsilon(s)}-\braket{\psi(s)}{A\psi(s)}| \leq C|\supp(A)|^2\|A\|\epsilon^k.
\]
where we recall that $k$ is related to the differentiability of the interaction. For $k=1$, the assumption that the derivative is compactly supported on $s\in(0,1)$ can be lifted. For more information on this result, Kubo's formula, and  the quantization of the Hall conduction, see \citep{Bachmann2018} and references therein.

\section{Stability of the spectral gap}
\label{sec:Stability}

In previous sections, we have discussed both uniformly gapped, frustration free quantum spin models, and the notion of a gapped ground state phase. These topics come together when studying stability of the spectral gap, where one tries to rigorously determine conditions when relatively general, small perturbations of a uniformly gapped quantum spin model belong to the same gapped phase.

\begin{figure}
    \centering
   \includegraphics[scale=.3]{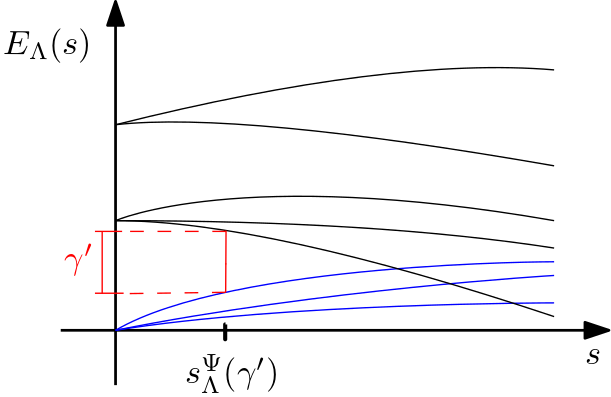}
    \caption{The eigenvalues of $H_\Lambda(s)$ as a function of $s$. The branches emanating from the origin correspond to the set $\Sigma_\Lambda^1(s).$}
    \label{fig:stability}
\end{figure}

To be more concrete, fix $\Lambda\in\cP_\Gamma$, and suppose that $H_\Lambda$ denotes the local Hamiltonian associated with an interaction $\Phi$, and $V_\Lambda$ is the local Hamiltonian associated with some other interaction $\Psi$, the latter of which plays the role of the perturbation. The eigenvalues of the perturbed Hamiltonian
\[
H_\Lambda(s) = H_\Lambda + s V_\Lambda, \quad s\in[0,1]
\]
are continuous functions of the parameter $s$ when written in ascending order: $E_\Lambda^{0}(s) \leq E_\Lambda^1(s) \leq \ldots E_\Lambda^N(s),$ where $N=\dim(\cH_\Lambda).$ Note that here we do not require the eigenvalues to be distinct, as compared to \eqref{distinct_eigenvalues}. Similar to the definition of a gapped ground state phase, the spectral gap of $H_\Lambda(s)$ is chosen to allow for potential eigenvalue splitting of the ground state energy, $\min\spec(H_\Lambda)$, see Figure~\ref{fig:stability}. To define this gap, first partition the spectrum of the perturbed Hamiltonian by:
\begin{align}
    \Sigma_\Lambda^1(s) & = \left\{E_\Lambda^{i}(s): E_\Lambda^{i}(0) =\min\spec(H_\Lambda)\right\}\label{gs_spec}\\
    \Sigma_\Lambda^2(s) & = \spec(H_\Lambda(s))\setminus \Sigma_\Lambda^1(s),\nonumber
\end{align}
and set $\gamma_\Lambda = \gap(H_\Lambda)$. The continuity of the eigenvalues guarantees that for any $0<\gamma'<\gamma_\Lambda$, there exists a maximal $s_\Lambda^\Psi(\gamma')\in(0,1]$ such that
\[
\gap(H_\Lambda(s)) := d\left(\Sigma_\Lambda^1(s),\Sigma_\Lambda^2(s)\right)\geq \gamma',\quad \forall\,  s \in[0,s_\Lambda^\Psi(\gamma')].
\]
If the unperturbed model is uniformly gapped, then there is $\Lambda_n\uparrow\Gamma$ so that $\gamma = \inf_n\gap(H_{\Lambda_n})>0$, and the spectral gap of the perturbed system is called ``uniformly stable'' if for all $0 < \gamma' < \gamma,$
\[
s^\Psi(\gamma') = \inf_n s_{\Lambda_n}^\Psi(\gamma') >0.
\]

In certain instances, one might be interested in a specific perturbation, $\Psi$. However, stability results typically aim to show that the spectral gap is stable for an entire class of perturbations with similar decay properties.  

As suggested at the beginning of this section, since methods for proving uniform gaps of quantum spin models are restricted to frustration free systems, the stability results to-date start from an assumption that the unperturbed interaction, $\Phi$, is frustration free. However, it is not required that the perturbed interaction, $\Phi+s\Psi$, also be frustration free. Hence, in addition to showing that a uniformly gapped interaction belongs to a nontrivial gapped ground state phase, gap stability also provides a path for proving uniform gaps for non-frustration free systems that are sufficiently small perturbations of frustration free, uniformly gapped systems.

A particularly powerful approach for gap stability, that utilizes the quasi-adiabatic continuation from Section~\ref{sec:gap_results}, was pioneered by Bravyi, Hastings, and Michalakis (BHM) for topologically ordered systems \citep{Bravyi2010}. One hallmark of such a system is that ground states cannot be distinguished by topologically trivial (i.e., local) observables. In terms of the finite volume ground states, this indistinguishability is captured by a property called local topological quantum order (LTQO). The BHM strategy yields uniform stability for a large class of perturbations of a uniformly gapped, frustration free model with LTQO.

For a fixed interaction $\Phi$, let $G_\Lambda$ denote the orthogonal projection onto the ground state space of a quantum spin Hamiltonian $H_\Lambda$, $\Lambda\in\cP_{\Gamma}$. A sufficient condition for $\Phi$ to have LTQO is the existence of a non-negative function, $\Omega$, decaying faster than any polynomial, and integer $p\geq 0$, such that for any $A\in\cA_{b_k(x)}$, $x\in\Gamma$ and $k\geq0$, there is a constant $c(A)$ so that for any $n>k$
\begin{equation}\label{LTQO}
    \|G_{b_n(x)}AG_{b_n(x)}-c(A)G_{b_n(x)}\| \leq  \|A\||b_k(n)|^p\Omega(n-k)\,.
\end{equation}
Here, $b_n(x)\subseteq \Gamma$ denote the ball of radius $n$ around $x\in\Gamma$.  More generally, one can allow for functions $\Omega$ with sufficient power law decay, but the exact conditions are a bit cumbersome. We should note that this LTQO condition is not the original one introduced by Bravyi, Hastings and Michalakis, where perturbations of CLP Hamiltonians were considered, but rather the criterion introduced in \citep{Michalakis2013} for the more general case of frustration free interactions.

Before returning to the stability result, note that the norm in \eqref{LTQO} captures that the projection of any local observable $A\in\cA_{b_k(x)}$ into the ground state space is approximately a constant, i.e. $c(A)G_{b_n(x)}$. Thus, the expectation of $A$ in any ground state will be $c(A)$ up to an error governed by $\Omega(n-k)$, and this becomes exact as $n\to\infty.$ As a result, any quantum spin model satisfying LTQO will have a unique, limiting ground state $\omega:\caA_\Gamma \to \bC$, and the constant in \eqref{LTQO} can be taken as the ground state expectation, $c(A)=\omega(A).$ If, in addition, the model is frustration free, it can be shown that the kernel of the GNS Hamiltonian associated to $\omega$ is one-dimensional, namely, $\ker(H) = \Span\{\Omega\}$ where $\Omega$ is as in \eqref{GNS}. 

 A spectral gap stability result that utilizes the BHM strategy for $\Gamma=\bZ^D$ is as follows. Suppose that the unperturbed interaction, $\Phi$, is finite range, frustration free, uniformly gapped, satisfies LTQO, and that the spectral gap of any local Hamiltonians supported on a ball does not decay too quickly; namely, there exists $\lambda>0$ and $q>0$ so that
 \begin{equation}\label{local_gap}
 \gap(H_{b_n(x)}) \geq \frac{\lambda}{n^q}, \quad \forall x\in\Gamma,\, n\geq 1.
 \end{equation}
 Then, the spectral gap is uniformly stable for any perturbation $\Psi$, such that $\|\Psi\|_{a,\theta}<\infty$, $a>0$ and $\theta\in(0,1]$. A similar statement can be made for more general $D$-dimensional lattices, so long as $|b_n(x)|$ does not grow too quickly in $n$.

 The proof of this result establishes gap stability for the unitarily transformed Hamiltonian  $\alpha_s^\Lambda(H_\Lambda(s))$ where $\alpha_s^\Lambda$ is the quasi-adiabatic continuation generated by the perturbed Hamiltonian $H_\Lambda(s) = H_\Lambda+sV_\Lambda$. Both the spectral flow property of the quasi-adiabatic continuation, \eqref{spectral_flow}, and the Lieb-Robinson bound of $\alpha_s^\Lambda$ are vital in this task. The end result shows that there is a constant $C_\Lambda(s)\in\bR$ and two observables $V_\Lambda(s),\Delta_\Lambda(s)\in\cA_\Lambda$ so that the transformed Hamiltonian can be written as
 \[
 \alpha_s^\Lambda(H_\Lambda(s)) = H_\Lambda+V_\Lambda(s)+\Delta_\Lambda(s) + C_\Lambda(s)\idty_\Lambda
 \]
 where $V_\Lambda(s)G_\Lambda = \Delta_\Lambda(s)(\idty_\Lambda-G_\Lambda)=0$, and there are positive constants $\beta_\Lambda$, $\epsilon_\Lambda$ with $\epsilon_\Lambda\downarrow 0$ as $\Lambda\uparrow\Gamma$ so that
 \begin{align}
 \|\Delta_\Lambda(s)\| & \leq s \epsilon_\Lambda\label{delta}\\ 
 \braket{\psi}{V_\Lambda(s)\psi} & \leq s\beta_\Lambda\braket{\psi}{H_\Lambda\psi}, \quad\forall\, \psi\in\cH_\Lambda.\label{form_bound}
 \end{align}
 
It is well known by perturbation theory that the form bound from \eqref{form_bound} implies the persistence of a gap, see \citep[Theorem 5.13]{Kato1995} or \citep[Lemma 3.2]{Nachtergaele2022}. Specifically, in the setting considered here,
\[
\spec(H_\Lambda+V_\Lambda(s)+\Delta_\Lambda(s))\cap (s\epsilon_\Lambda,(1-s\beta_\Lambda)\gamma_\Lambda-s\epsilon_\Lambda) = \emptyset
\]
where $\gamma_\Lambda = \gap(H_\Lambda)$.  The uniform stability of the gap is a result of showing that $\beta:=\sup_n\beta_{\Lambda_n}$ is finite, where $\Lambda_n\uparrow\bZ^D$ is the IAS for the uniform gap of the unperturbed model. The norm $\|\Delta_\Lambda(s)\|$ controls the eigenvalue splitting around the ground state energy. Specifically, \eqref{delta} implies that the set from \eqref{gs_spec} is contained in the interval
\[
\Sigma_\Lambda^1(s) \subseteq \big[C_\Lambda(s)-s\epsilon_\Lambda, C_\Lambda(s)+s\epsilon_\Lambda\big].
\] 

The BHM strategy for gap stability has been generalized to deal with a number of different settings than what has been presented here. The LTQO condition can be modified to establish gap stability for models with discrete symmetry breaking. It can also be applied to prove gap stability of the GNS Hamiltonians along the path $\omega_s=\omega\circ \alpha_s$, where $\omega$ is the unique limiting ground state implied by the LTQO property. Moreover, one can lift the uniform gap requirement so long as the GNS Hamiltonian is gapped and the local gap condition \eqref{local_gap} still holds. Proofs of the discrete symmetry breaking and infinite volume results can be found in \citep{Nachtergaele2021, Nachtergaele2022}.

While we have focused on the BHM stability strategy, this is not the only approach to stability. For specific models, stability has also been proved using cluster expansions. There are also stability results for weakly interaction lattice fermion models, including one that adapts the BHM strategy. Another approach utilizes a Lie-Schwinger diagonalization scheme. This approach can also be applied to certain classes of lattice boson models whose interactions are unbounded on-site terms. However, the BHM strategy is the one that applies to the largest class of models, which is why it is the focus here. One commonality amongst all of these approaches is that the gap stability is ultimately a result of a bound similar to \eqref{form_bound}. For more information on the BHM approach, as well as references to the other stability methods discussed above, see \citep{Nachtergaele2022}.

\section*{Further Readings} In addition to the references made in the text, more information on the foundations of quantum spin systems can be found \citep{Bratteli1987,bratteli:1997_2}. For a text including more recent topics, including topological phases, see \citep{Tasaki2020}. For information and references on different spectral gap techniques, see \citep{Young2023}.

\section*{Acknowledgements} This work was funded by the DFG under TRR 352 (Grant number 470903074):  Mathematik der Vielteilchen-Quantensysteme und ihrer kollektiven Phänomene .

\bibliographystyle{elsarticle-harv} 
\bibliography{EMP.bib}






\end{document}